\documentstyle[prb,aps,epsfig,twocolumn]{revtex}
\begin{document}
\draft
\flushbottom
\twocolumn[
\hsize\textwidth\columnwidth\hsize\csname @twocolumnfalse\endcsname

\author{R. Narevich, R. E. Prange and Oleg Zaitsev}
\address{Department of Physics, University of Maryland, College Park, MD 20742}
\title{Localization by interference: Square billiard with a magnetic flux}
\date{\today}
\maketitle

\tightenlines
\widetext
\advance\leftskip by 57pt
\advance\rightskip by 57pt
\begin{abstract}
Eigenstates and energy levels of a square quantum billiard in a magnetic
field, or with an Aharonov-Bohm flux line, are found in quasiclassical
approximation, that is, for high enough energy. Explicit formulas for the
energy levels and wavefunctions are found. There are localized states, never
before noticed in this well studied problem, whose localization is due to
phase interference, even though there is no or negligible classical effect
of the magnetic field. These and related states account almost entirely for
the magnetic response in certain temperature ranges, and thus have a bearing
on the experiments of L\'evy, {\em et. al.}\cite{Levy}
\end{abstract}


]

\narrowtext
\tightenlines
The problem of a quantum charged particle subject to a magnetic flux and
confined in two dimensions, say to a square billiard, has attracted much
attention. The simplest fluxes are (1) a uniform magnetic field [UF] and (2)
an Aharonov-Bohm flux line [ABFL]. One motivation for (1) is experimental
mesoscopic physics. L\'evy, {\em et al}$,$\cite{Levy} have made
theoretically resistant\cite{Richter} measurements of the magnetic
susceptibility of an array of such squares. Clearly, such a simple problem
should be added to the physicist's arsenal of known results.

The Aharonov-Bohm flux line\cite{abfl} has had a rather spectacular effect
on physics. It provides a compelling example of quantum nonlocality. Largely
because of this example, it is now clearly understood that charged quantum
particles have phase interference effects originating from a magnetic field
that vanishes in all regions accessible to the particle. Although this
purely quantum effect has no directly corresponding classical physics, it is
related to the scattering of a classical charge neutral wave from a vortex%
\cite{berryv}. Aside from this conceptual motivation, an ABFL is a simple
idealized extreme case, which like the uniform field should be thoroughly
understood as there are numerous applications. For example, Laughlin's
explanation\cite{Laughlin} of the quantum Hall effect is based on ABFL
physics.

Another reason for interest is that a magnetic flux is the most obvious way
to break time reversal symmetry of simple systems like billiards\cite
{universal}. The finite flux case is in a different universality class from
the zero flux case. This has inspired much numerical work\cite{Date}. Also,
square symmetry conflicts with the apparent circular symmetry of the flux.
This, as for the Sinai billiard, suggests chaos\cite{chaos}. Indeed, the
circular billiard in a uniform field is relatively trivial and uninteresting.

Thus, much work has been devoted to these systems. It is surprising that
there is something new, simple and interesting to say about square billiards
with a flux.

We here present some analytic results with numerical confirmation. The
results are possible because we exploit two small parameters. First, we
treat the particle {\em quasiclassically,} that is, its wavelength is short
compared with the system parameters. Second, we treat the field as {\em %
classically weak, }which is automatic for the ABFL. For the UF, we assume
that $\epsilon =L/R_c$ is small, where $L$ is the square side and $R_c$ is
the cyclotron radius.

Nearly all previous work has focused on level statistics, or smoothed state
densities. In contrast, we find{\em \ }analytic {\em wave functions, }and
simple formulas for their energies. The {\em unexpected result} is that an
important class of these eigenstates are {\em localized.} Unlike the
eigenstates of the flux-free case, which uniformly fill the square, some of
the eigenstates have support in a small fraction of the square. The area of
this support vanishes at infinite energy. Associated with a given extremely
localized state is a sequence of states which progressively become more
delocalized, but which share with the localized state a simple structure.
The most prominent such sequence {\em dominates }the magnetic susceptibility
in the UF case, as measured by L\'evy.

This localization, like Anderson localization\cite{anderson}, is caused by 
{\em phase interference}. A bunch of classical phase space points mimicking
a wave packet in a disordered two dimensional region, or in a square, will
spread, diffusively in the disordered case, linearly in the square, and
eventually become uniformly distributed. Taking into account the phases in
impurity scattering, or from the magnetic flux, leads to interference
effects that suppress the wave packet spreading. Many interesting effects of
the ABFL have been discovered in the last forty years, but we have found no
previous work that produces a localized quantum state through destructive
ABFL interference.

In this paper we shall not discuss to its conclusion the ideal zero radius
ABFL. That limit best establishes quantum nonlocality, but quantum
nonlocality is no longer disputed. We have found that the zero radius limit
leads to interesting but distracting mathematical and numerical problems.
Basically, there are diffraction effects which complicate the results. In
this paper we want to avoid that issue in favor of understanding the main
phase interference effect. We therefore eventually endow the flux line with
a finite radius $\rho $ whose scale we discuss later. Initially, however, we
suppose $\rho =0.$

For UF, the cyclotron radius is $R_c=v/\omega _c=cp/eB,$ where $\omega
_c=eB/mc$ and $p=mv.$ The momentum $p=\hbar k=h/\lambda $ is related to
wavenumber $k$ and wavelength $\lambda $. Then $\epsilon =L/R_c=eBL/\,\hbar
ck=2\pi \phi /\phi _0kL$. Here $\phi $ is the magnetic flux $BL^2,$ and $%
\phi _0\,$is the flux quantum $hc/e.$ This expression carries over to the
ABFL if we identify $\phi $ with the flux in the flux line$.$ For the ABFL
we may as well take $\left| \phi /\phi _0\right| \leq 0.5,\ $but for the UF $%
\phi $ can be large as long as $\epsilon $ is small. We choose units such
that $B\equiv 2\pi \phi /\phi _0$ and $L,\,\hbar $ and $2m\ $are unity, so
that $\epsilon =B/k.$ We also suppose $k>>1,$ justifying the quasiclassical
approximation.

Our approach\cite{pnz} utilizes the quasiclassical surface of section [SS]
method of Bogomolny\cite{bogolss}. His operator $T(x,x^{\prime };E)$ takes
the electron crossing the SS at point $x^{\prime }$ to its next crossing at $%
x,$ all at energy $E=k^2.\,$ The SS can be chosen in many ways. To simplify $%
T$, we use a method of images. Namely, we consider, instead of a unit
square, $x,y\in [0,1]\otimes [0,1],$ an infinite channel of width $2$
obtained by reflecting the original square first about $x=0$ and then about $%
y=1,$ and finally repeating the resulting $2\times 2$ square periodically to 
$\left| x\right| =\infty $. The flux changes sign in neighboring squares.
The solutions to the square are found from the channel solutions by using
symmetry, {\em e.g.}, the solution odd under $y\leftrightarrow 2-y$
corresponds to Dirichlet conditions at $y=1.$

The SS is taken as the axis $y=0\,$ which is identified with $y=2.$ This
gives\cite{bogolss} 
\begin{equation}
T(x,x^{\prime };E)=\left( \frac 1{2\pi i}\left| \frac{\partial
^2S(x,x^{\prime };E)}{\partial x\partial x^{\prime }}\right| \right) ^{\frac 
12}\exp \left( iS(x,x^{\prime };E)\right) .  \label{T}
\end{equation}
Here $S=\int_{x^{\prime }}^x{\bf p}\cdot d{\bf r}\,${\bf \ }is the action
integral along the classical path from ($x^{\prime },$ $0$) to ($x,$ $2$).
Because the field is classically weak, this path is approximated by a
straight line. We immediately find 
\begin{equation}
S(x,x^{\prime })=k\sqrt{4+\left( x-x^{\prime }\right) ^2}+\Phi (x,x^{\prime
}),  \label{S}
\end{equation}
the flux free result plus $\Phi =(e/c)\int {\bf A\cdot }d{\bf r}$.

Periodic orbits on the square correspond to straight line orbits in the
channel from ($x^{\prime },0$) to ($x=x^{\prime }+2p/q,$ $2$). Here $q$ is a
positive integer and $p$ is a positive or negative integer relatively prime
to $q.$ Negative and positive $p$ are not equivalent if there is a magnetic
flux. Such orbits correspond to a ($p,q$) classical resonance which is
strongly affected by a perturbation.

Our scheme finds solutions of $T\psi =\psi $ by a perturbation theory\cite
{pnz}. We first solve $\int T(x,x^{\prime })\psi (x^{\prime })=e^{i\omega
(k)}\psi (x)$, treating $k$ as a parameter, and then find the energies by
solving $\omega (k)=2\pi n.$ Given $\psi (x),$ a quadrature yields the full
wave function $\Psi (x,y)$.$\,$ Given $\Psi ,$ for billiards $\psi
(x)\propto \partial \Psi /\partial n\propto \partial \Psi (x,y)/\partial
y|_{y=0}.$

Specializing to the (1,1) resonance we look for a solution $\psi
(x)=e^{i\kappa x}u_m(x-\frac 12),$ where $\kappa =k\cos 45^{\circ },$ and $%
u_m$ varies much more slowly than the exponential. The rapidly varying
phases in the integral $\int T\psi $ are stationary at $x^{\prime }=x-2.$
This corresponds to diamond shaped periodic orbits of the original square
whose sides make angles of $45^{\circ }$ with the $x$ axis. It suffices\cite
{pnz} to evaluate $\Phi (x,x^{\prime })$ at $\Phi (x,x-2)=\Phi (x+2,x).$
This is equivalent to integrating the vector potential about the closed
diamond loop, and so is independent of gauge. The result is that $u_m$
satisfies the Schr\"odinger equation 
\begin{equation}
-u_m^{\prime \prime }+V(x)u_m=E_mu_m,  \label{um}
\end{equation}
where $V(x-\frac 12)=-k\Phi (x+2,x)/{\cal L},$ and ${\cal L}=\sqrt{8}$ is
the length of the diamond orbits. Thus we convert the phase $\Phi $ to a
`potential' $V.$ For UF, 
\begin{eqnarray}
V(x) &=&-Bk(%
{\textstyle {1 \over 2}}
-2x^2)/{\cal L};\;\;x\in [-%
{\textstyle {1 \over 2}}
,%
{\textstyle {1 \over 2}}
],  \nonumber \\
V(x) &=&+Bk(%
{\textstyle {1 \over 2}}
-2(x+1)^2)/{\cal L};\;\;x\in [-%
{\textstyle {3 \over 2}}
,-%
{\textstyle {1 \over 2}}
],  \nonumber \\
V(x) &=&V(x+2).  \label{Vuf}
\end{eqnarray}

For the (-1,1) resonance, whose orbits are time reversed (1,1) orbits, $V(x)$
changes sign. This would not be true if $V$ had its origin in a time
reversal invariant perturbation of the square, for example, a small change
of shape. Note also that this periodic extension of the $x$ coordinate is
similar to use of a `angle' variable, with positive $x$-velocity $v_x$ for $%
x\in [0,1],$ and negative $v_x$ for $x\in [1,2].$

For ABFL, we must choose where to put the flux line. The square center
offers some numerical convenience\cite{Date}, but the results are not very
interesting. We place the line at ($\frac 12,a$) which still gives some
symmetry and is readily compared with the UF case. Then 
\begin{eqnarray}
V(x) &=&-Bk/{\cal L};\;x\in [-a,a],  \nonumber \\
V(x) &=&+Bk/{\cal L};\;x\in [-1-a,-1+a],  \label{Vabfl}
\end{eqnarray}
and for other points in $[-\frac 32,\frac 12],$ $V(x)=0.$ This `square well'
potential is also continued periodically. Strictly speaking, this potential,
because of the steps, varies too rapidly for the theory to apply. We will
modify it below.

It's clear that $Bk$ is the important parameter in the UF case, while for
the ABFL, both $Bk$ and $a$ are important. For sufficiently large $Bk$, Eq. (%
\ref{um}), in the UF case, will have low lying tight binding harmonic
oscillator type solutions centered at $x=0$ (if $B>0$),\thinspace with
energies 
\begin{equation}
E_m=-%
{\textstyle {1 \over 2}}
Bk/{\cal L}+(m+%
{\textstyle {1 \over 2}}
)\sqrt{8Bk/{\cal L}}.  \label{EmU}
\end{equation}
The lowest wave function is approximately $u_0(x)=e^{-\sqrt{Bk/2{\cal L}}%
x^2} $ which is arbitrarily narrow at large energy.

For the ABFL 
\begin{eqnarray}
u_{m-1}(x) &\approx &\sin \left( m\pi (x+a)/(2a)\right) ;\;\;\;\left|
x\right| <a  \nonumber  \label{HO} \\
\ &\approx &0;\;\;\;\;\;\;\;\;\left| x\right| >a  \label{sw}
\end{eqnarray}
and the energy is 
\begin{equation}
E_m\approx -Bk/{\cal L}+(m+1)^2\pi ^2/4a^2.  \label{Em}
\end{equation}
Here the width of the state $u$ is determined by $a,$ and the tight binding
approximation will be good provided $Ba^2k/{\cal L}>>1.$ The localization of
these states in $x$ leads to the localization of the full state in the
square.

The total energy is given approximately by\cite{pnz} 
\begin{equation}
k_{n,m}=2\pi n/{\cal L+}E_m/k.  \label{knm}
\end{equation}
Eq. (\ref{knm}) should be solved iteratively. For example, the first
approximation replaces the $k$ dependence of the term $E_m/k$ by $2\pi n/%
{\cal L}.$ Equivalently, the energy $E_{n,m}=4\pi ^2n^2/{\cal L}^2+2E_m.$

We next give an expression for $\Psi (x,y),$ shifting the origin to the
square center: 
\begin{equation}
\Psi _{nm}(x,y)=\left( \sum_{s=0}^3i^{-rs}{\cal R}^s\right) e^{i\frac \pi 2%
n(x+y)}u_m(x-y-%
{\textstyle {1 \over 2}}
).  \label{Psi}
\end{equation}
Here ${\cal R}:(x,y)\rightarrow (y,-x)$ is the rotation by $90^{\circ }.$
The UF Hamiltonian can be taken to be invariant under ${\cal R}$, while the
approximations made for the ABFL induces this invariance. Here $r$ labels
the rotational symmetry of the wave function, i.e. ${\cal R}=i^r.$ Another
symmetry gives $u_m(x)=(-1)^mu_m(-x)$. The symmetry under translation is $%
u(x+2)=(-1)^ru(x).$ The relation is 
\begin{equation}
r=-n%
\mathop{\rm mod}
4+2(1-m%
\mathop{\rm mod}
2).  \label{groupr}
\end{equation}
Thus Eq. (\ref{knm})\thinspace holds for all symmetries and successive
values of $n$ cycle through the representations of ${\cal R}$.

It is easily checked that the $\Psi _{nm}$ vanishes on the square sides.
This wave function can also be obtained as a kind of Born-Oppenheimer or
channelling approximation\cite{BO}. To see that $\psi _m$ $\approx $ $%
\partial \Psi /\partial n,$ just observe that the derivative, say $\partial
/\partial y$ at $y=-\frac 12,$ need be applied only to the rapidly varying
exponential.

Fig. 1. shows numerically calculated $\left| \Psi (x,y)\right| ,\,$ which is 
gauge invariant, for $%
n=62,$ $k\approx 2\pi 62/{\cal L}\approx 138,$ $m=0,$ $B=25,$ $\sqrt{Bk}%
\approx 59.$ Current density for the same state is shown in Fig. 2. For
such a well localized $u_m$, each term in Eq.(\ref{Psi})
dominates one side of the diamond orbit. In this case $\left| \psi _m\right|
\approx $ $u_m.$ There is interference near the square's edges, so the first
maximum of $\left| \Psi \right| $ ({\em e.g. }near ($0,\pm \frac 12$)) is
about twice as large as $\left| \Psi (\pm \frac 14,\pm \frac 14)\right| $.
The current density is thus largest close to the middle of the square's
edges. These states are paramagnetic, that is, their current circulates in
the opposite sense from that of a free particle in the field.

There is no localized state in the time reversed sense. However, the states
with $m$ of order $\sqrt{Bk/{\cal L}^3}$, corresponding to $E_m\sim +Bk/2%
{\cal L}$, energetically at the top of `potential' $V(x),$ are diamagnetic.
This actually corresponds to the (1,-1) resonance. These states are not
spatially localized, although they do have a sort of localization in
momentum space, as we show in the next paragraph. More generally, as $m$
increases, the states become more delocalized, and eventually become
independent of $B.$ This means that for larger $m$ all four terms in Eq. (%
\ref{Psi}) make comparable contributions at an arbitrary typical point $x,y$%
, whereas in the localized case, only one or two terms contribute. This
gives interference oscillations in $\left| \psi _m\right| $ near $\left|
x\right| \approx \frac 12$ as shown in Fig. 3b).

We have assumed that $n>>m,$ and that $u_m$ is slowly varying compared with $%
e^{i\pi nx/2}.$ Expanding, $u_m=\sum \hat u_{m,l}e^{i\pi lx}$, where $2l$ is
an integer satisfying $(-1)^{2l}=(-1)^r.$   Also $\hat u_{m,l}=(-1)^m\hat u%
_{m,-l}.$ The unperturbed states can be labelled $p,q$ with unperturbed
energies $(p^2+q^2)$, dropping the factor $\pi ^2.$ For $r$ even, Eq. (\ref
{Psi}) is a superposition of unperturbed states with quantum numbers $p=%
\frac 12n+l,$ $q=\frac 12n-l,$ where $l<<\frac 12n.\,$ For $r$ odd, $%
p=n^{\prime }+l^{\prime }+1,$ $q=n^{\prime }-l^{\prime },$ $n^{\prime
},l^{\prime }$ integer. The even $r$ states have unperturbed energies $\frac 
12n^2+2l^2$ and so are nearly `degenerate' to the base energy $\epsilon _n=%
\frac 12n^2$. In particular, they are closer to $\epsilon _n$ than to the
base energy of the next representation, $\epsilon _{n\pm 1}\approx \epsilon
_n\pm n.$ If the perturbation is symmetric under rotation, the next base
energies coupled are $\epsilon _{n\pm 4}\approx \epsilon _n\pm 4n.$ There
are, however, many unperturbed states with $p^2+q^2\approx \epsilon _n.$ For
example, $7^2+49^2=\epsilon _{70}.$ However, the matrix elements ${\cal H}%
_{pq,p^{\prime }q^{\prime }}$ of a smoothly perturbed Hamiltonian, in the
unperturbed basis, are small if $\left| p-p^{\prime }\right| $ or $\left|
q-q^{\prime }\right| $ is large. Thus, an interpretation of our method,
which yields the states of Eq. (\ref{Psi}), is that we effectively
diagonalize the Hamiltonian in a basis restricted to the unperturbed states
`degenerate' with $\epsilon _n$ and close to $\frac 12n,\;\frac 12n.$ This
is the case for the UF, and indeed, we achieve agreement between full
numerical diagonalization, diagonalization restricted to `degenerate'
states, and the procedure using the solution of the differential equation
Eq. (\ref{um}).

\begin{figure}[tpb]
{\hspace*{-0.0cm}\psfig{figure=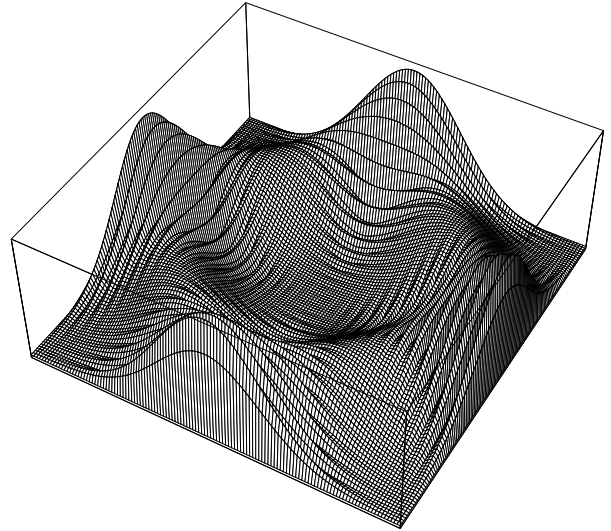,height=7cm,angle=0}} 
{\vspace*{.13in}}
\caption[ty]{Absolute value of a localized wave function on the unit square. Uniform field, B=25, n=62, m=0.}
\label{fig:1}
\end{figure} 

For the ideal, zero radius ABFL, there are significant deviations from this
scenario. Indeed, most matrix elements of the ABFL perturbed Hamiltonian in
the unperturbed basis are infinite. However, it is a weak, logarithmic
infinity, and our theory seems to capture the main shape of the wave
function, although at the relatively low energies, for which numerical
results are available, there are significant corrections. We consider these
to be diffraction corrections, arising from a characteristic length shorter
than the wavelength.

We therefore use a finite radius, $\rho $, for the flux line. The field
inside the flux tube is $B_0=\phi /\pi \rho ^2.$ The typical angular
deflection suffered by a particle traversing this field is $\delta \theta
\sim (\phi /\phi _0)/k\rho .$ To avoid diffraction we require $\delta \theta 
$ to be small. In the numerical work shown, we take $\phi /\phi _0=0.1,$ and 
$\rho =0.01,$ while $k\geq 140.$ This hardly changes the effective potential
of Eq. (\ref{Vabfl}). An alternative and equivalent condition is to insist
that, on the appropriately defined average\cite{BO}, the terms in the
Hamiltonian satisfy $\left\langle (eA/c)^2/2m\right\rangle $ $<<\left\langle
e{\bf p\cdot A}/mc\right\rangle $.

Fig. 3 is for the ABFL case, with $n=58,$ $m=0,$ $r=0$ and $a=\frac 14.$
Fig. 3a) shows $u_0(x)\ $ and $u_0(-x-1)$, its extension into $x<-\frac 12,$
reflected. For these parameters, $u_0$ is not extremely localized, and
extends significantly outside $[-\frac 12,\frac 12].$ The remaining plots
give $\left| \psi _m\right| =\left| \partial \Psi /\partial n\right| .$ Fig.
3b) plots Eq. ( \ref{Psi}), 3c) is from diagonalization in the limited basis
of the `degenerate' states; and 3d) is from numerical diagonalization in the
complete unperturbed basis. To give the flavor of the diffraction effects,
we show two numerical results for $\left| \psi _m\right| $ for a zero radius
flux line obtained by a special numerical method\cite{BO}: Fig. 3e) uses the
above parameters, and 3f) is for $n=70.$

We finally argue that the (1,1) states dominate the orbital susceptibility
in a parameter range appropriate to experiments. The orbital susceptibility $%
\chi $ of a system of noninteracting spinless electrons is given by $\chi
=\partial {\cal M}/\partial B$ where the magnetization ${\cal M}=-\partial
\Omega (T,\mu ,B)/\partial B$ is that of the grand canonical ensemble, 
\begin{equation}
{\cal M}(T,\mu ,B)=\sum_{n,m}\frac{\partial E_{n,m}}{\partial B}%
f_D(E_{n,m}(B)).  \label{Gr}
\end{equation}
where $f_D$ is the Fermi-Dirac distribution function. [The canonical
ensemble is required in averaging over many such billiards, but for a single
billiard the grand canonical suffices\cite{Richter}.] The chemical potential 
$\mu =k_F^2$ is nearly independent of $B$, since the many states not
depending much on field act as a heat bath.

\begin{figure}[tpb]
{\hspace*{0.cm}\psfig{figure=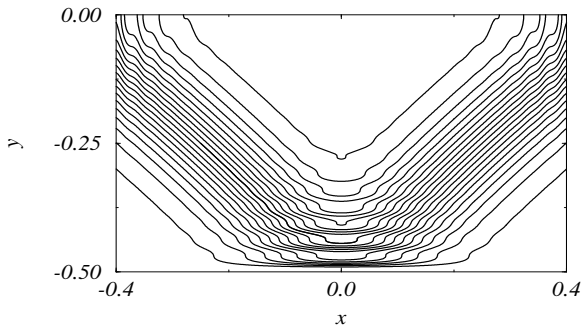,width=4.5cm,height=8cm,angle=270}}
{\vspace*{.23in}}
\caption[ty]{Current density of the wave function shown in Fig. 1. The
stream lines are plotted, their density proportional to the current
density. The sense of the current is paramagnetic.}
\label{fig:3}
\end{figure}                                                                  

It is easiest, no doubt, to estimate Eq. (\ref{Gr}) using the perturbed
Berry-Tabor trace formula\cite{Richter}. However, it can be shown that using
Eq. (\ref{knm}) in Eq. (\ref{Gr}) above is equivalent to keeping the (1, $%
\pm $1) periodic orbits and their repetitions in the trace formula. Using
the Poisson sum formula, replace the sum on $n$ in Eq. (\ref{Gr}) by an
integral over $k$, and do the integral to obtain 
\begin{equation}
{\cal M}=\sum_{r,m,s=0}^\infty \frac{2{\bf \alpha }_m{\cal L}\omega _0}{\pi
k_F}\exp \left( -\frac{\omega _rs{\cal L}}{2k_F}\right) \sin ({\cal L}s(k_F-%
\frac{E_m}{k_F})).  \label{M}
\end{equation}
Here, $\omega _r=\pi (2r+1)k_BT$ and $\alpha _m=\partial E_m/\partial B.$ As
an example, take $k_BT$ ten times the level spacing $\bar d^{-1}$ of all
levels, i.e. $k_BT=40\pi .$ Then, $\omega _0{\cal L}/2\approx 560.$ If $%
k_F\approx 500,$ so that the square contains about 4000 electrons, the
exponential suppression will not be too serious for $r=0,$ $s=1$. Doing the $%
m$ sum is possible but harder\cite{BO}.

Eq. (\ref{M}) is obtained from explicit energy levels rather than the action
of periodic orbits as found from the trace formula, but the result is the
same. Integrable systems have regularities in their energy levels which lead
to larger effects in quantities like ${\cal M}$ as compared with chaotic
systems. For the square, the (1,1) states have the smallest ${\cal L}$ and
also the largest $\alpha _m$. The (2,1) resonance does not couple to a
constant field. The (3,1) resonances have length ${\cal L}_{31}=$ $2\sqrt{10}
$ $=\sqrt{5}{\cal L}_{11}$ and a potential $V_{31}(x)=V_{11}(x)/3\sqrt{5}$.
An interesting case is the (1,0) resonance which has ${\cal L}_{10}=2$, but
this resonance corresponds to classical orbits enclosing no flux. At
stronger fields, the curvature of the orbits must be allowed for, and
eventually this resonance dominates\cite{Richter}. At very strong fields,
such that $R_c<L,$ the standard de Haas-van Alphen susceptibility is
recovered.

\begin{figure}[tpb]
{\hspace*{0.2cm}\psfig{figure=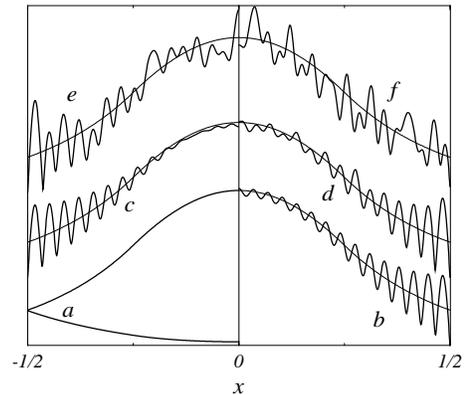,width=6cm,height=7.5cm,angle=270}} 
{\vspace*{.23in}}
\caption[ty]{ABFL case. a)-f) $u_0,$ and b)-f) $\left| \partial \Psi /\partial n\right| ,$ as a function of $x,$ for several cases described in the text.}
\label{fig:2}
\end{figure}

In conclusion, a new technique\cite{pnz} allows analytic solution of problems 
of integrable systems subjected to magnetic flux. Solvable problems include 
billiards which are nearly rectangular, hexagonal or elliptical, as well as
coupled nonlinear oscillators, with uniform or
nonuniform magnetic flux. Some of the solutions are
striking, unexpected and experimentally relevant. Supported in part by NSF
DMR 9624559 and U. S.-Israel BSF 95-00067-2. REP thanks ITP Santa Barbara
for support and hospitality during the early phases of the work.

\end{document}